\documentclass[%
reprint,
 amsmath,amssymb,
 aps,
prd,
floatfix,
]{revtex4-1}

\usepackage{tikz}
\usetikzlibrary{decorations.pathreplacing}
 \usetikzlibrary{plotmarks}
\usepackage{subfigure}
\usepackage{booktabs}
\usepackage{graphicx}
\usepackage{dcolumn}
\usepackage{bm}
\usepackage{youngtab}
\usepackage{ytableau}

\usepackage{braket}
\usepackage{multirow}
\usepackage{dsfont}
\usepackage{mathtools}
\usepackage{xcolor}
\usepackage[normalem]{ulem}

\begin{document}

\title{Baryon parity doublets and chiralspin symmetry}

\author{M.~Catillo}
\author{L.~Ya.~Glozman}

\affiliation{Institute of Physics, University of Graz, 8010 Graz, Austria}
 
\date{\today}

\begin{abstract}
The chirally symmetric baryon parity-doublet model can be used as an effective
description of the baryon-like objects in the chirally symmetric
phase of QCD. Recently it has been found
that above the critical temperature higher chiralspin symmetries emerge in QCD.
It is demonstrated here that the baryon parity-doublet Lagrangian is manifestly
chiralspin-invariant. We construct  nucleon interpolators with fixed chiralspin transformation properties that can be used in lattice studies at high T.
\end{abstract}
\maketitle

\section{\label{sec:intro}Baryon parity doublets. Introduction}

A Dirac Lagrangian of a  massless fermion field  
is chirally symmetric since the left- and right-handed components
of the fermion field are decoupled,

\begin{equation}
{\cal L} = i \bar \psi \gamma_\mu \partial^\mu \psi =
i \bar \psi_L \gamma_\mu \partial^\mu \psi_L +
i \bar \psi_R \gamma_\mu \partial^\mu \psi_R,
\label{L0}
\end{equation}
where 
\begin{equation}
\psi_R = \frac{1}{2}\left( 1+\gamma_5 \right ) \psi,~~
\psi_L = \frac{1}{2}\left( 1-\gamma_5 \right ) \psi.
\label{RL}
\end{equation}

\noindent
The $SU(2)_L \times SU(2)_R$ chiral
symmetry is a symmetry upon independent isospin rotations of the
left- and right-handed components of fermions, 

\begin{equation}
\psi_R \rightarrow 
\exp \left( \imath \frac{\theta^a_R \tau^a}{2}\right)\psi_R; ~~
\psi_L \rightarrow 
\exp \left( \imath \frac{\theta^a_L\tau^a}{2}\right)\psi_L,
\label{ROT}
\end{equation}

\noindent
where $\tau^a$ are the isospin Pauli matrices
  and the
angles $\theta^a_R$ and $\theta^a_L$ parameterize rotations
of the right- and left-handed components. The transformation
(\ref{ROT}) defines the  $(0,1/2) \oplus (1/2,0)$ representation 
of the chiral group.

The mass term in the
Lagrangian breaks explicitly the chiral symmetry since it couples the
left- and right-handed fermions.

It is known for a long time that it is possible to construct
a chirally symmetric Lagrangian for a massive fermion field provided
that there are mass-degenerate fermions of opposite parity - parity doublets -
that transform into each other upon a chiral transformation \cite{LEE}.

Consider a pair of the isodoublet fermion fields 

\begin{equation}
\Psi = \left(\begin{array}{c}
\Psi_+\\
\Psi_-
\end{array} \right),
\label{doub}
\end{equation}

\noindent
where the Dirac bispinors
$\Psi_+$ and $\Psi_-$ have positive and negative parity, respectively.
The parity doublet above is a spinor constructed from
two Dirac bispinors and contains eight components. Note that there
is in addition an isospin index which is suppressed.
Given that the right- and left-handed fields are
directly connected with the opposite parity fields

\begin{equation}
\Psi_R = \frac{1}{\sqrt{2}}\left( \Psi_+ + \Psi_-\right);~~
 \Psi_L = \frac{1}{\sqrt{2}}\left( \Psi_+ - \Psi_-\right),
\label{RLL}
\end{equation}

\noindent
the vectorial and axial parts of the chiral transformation law
 under the $(0,1/2) \oplus (1/2,0)$ representation 
of $SU(2)_L \times SU(2)_R$ is

\begin{equation}
\Psi \rightarrow 
\exp \left( \imath \frac{\theta^a_V \tau^a}{2}
\otimes \mathds{1}\right)\Psi; ~~
\Psi \rightarrow 
\exp \left(  \imath \frac{\theta^a_A\tau^a}{2}\otimes \sigma_1
\right)\Psi.
\label{VAD}
\end{equation}

\noindent
Here $\sigma_i$ is a Pauli matrix that acts in the  
space of the parity doublet. In the  chiral transformation
law (\ref{ROT}) the axial rotation mixes the massless Dirac spinor $\psi$
with $\gamma_5 \psi$ , while in the chiral rotation of the parity doublet  
 a mixing of two independent fields $\Psi_+$ and $\Psi_-$ is provided.
Then the chiral-invariant Lagrangian of the free parity doublet is given as

\begin{equation}
\begin{split}
\mathcal{L} & = i \bar{\Psi} \gamma^\mu \partial_\mu \Psi - m \bar{\Psi}
\Psi  \\
& =  i \bar{\Psi}_+ \gamma^\mu \partial_\mu \Psi_+ + 
i \bar{\Psi}_- \gamma^\mu \partial_\mu \Psi_-
- m \bar{\Psi}_+ \Psi_+ - m \bar{\Psi}_- \Psi_-.  \\
\end{split}
\label{lag}
\end{equation}

\noindent
Note that this Lagrangian can be written in different
equivalent forms \cite{DETAR,TIT}. The equivalence of the present
form  \cite{LEE} with that one of ref. \cite{TIT} was demonstrated in ref. 
\cite{G1} and the chiral transformation law (\ref{VAD}) corresponds
to the ``mirror" assignment of ref. \cite{TIT}. 
For the reader convenience we derive this equivalence in
Appendix \ref{app:A}.

This Lagrangian can also be equivalently written
in terms of the right- and left-handed fields of eq. (\ref{RLL}),

\begin{equation}
\begin{split}
\mathcal{L} 
& =  i \bar{\Psi}_L \gamma^\mu \partial_\mu \Psi_L + 
i \bar{\Psi}_R \gamma^\mu \partial_\mu \Psi_R
- m \bar{\Psi}_L \Psi_L - m \bar{\Psi}_R \Psi_R.
\end{split}
\label{llag}
\end{equation}

\noindent
The latter form demonstrates  that the right-
and left-handed degrees of freedom are completely
decoupled and the Lagrangian is manifestly chiral-invariant.
Since the latter Lagrangian is equivalent to the
Lagrangian (\ref{lag}) it is also manifestly Lorentz-invariant (in order to avoid confusion note that $\Psi_L$ and $\Psi_R$ for the parity doublet are defined differently from the massless Dirac bispinor (2)).

 A crucial
property of the  Lagrangian (\ref{lag}-\ref{llag}) is that  the fermions
$\Psi_+$ and $\Psi_-$ are exactly degenerate and
have a nonzero chiral-invariant mass $m$.
This Lagrangian has a number of peculiarities. In particular, the diagonal
axial charge of the fermions $\Psi_-$ and $\Psi_+$ vanishes, while the off-diagonal
axial charge is 1.

Supplemented by the interactions with the pion and
sigma-fields \cite{DETAR,TIT} this Lagrangian has been sometimes used 
in baryon spectroscopy
\cite{Rischke} and as  an effective model for chiral symmetry restoration scenario at high temperature or density  where  baryons
with the nonzero mass do not vanish upon a chiral restoration \cite{Sasaki}.

Recently it was found on the lattice \cite{R} that at temperatures above
the critical one  new symmetries emerge in QCD, $SU(2)_{CS}$
and $SU(2N_F)$ \cite{Glozman:2014mka, Glozman:2015qva}, 
which implies a modification of the existing view
on the nature of the strongly interacting matter at high temperatures
\cite{G2}. In particular, the elementary objects at high temperatures are
 quarks with a definite chirality connected by the chromo-electric field, i.e. there are no
 free deconfined quarks.

 These symmetries were observed earlier 
upon artificial truncation  of the near-zero
modes of the Dirac operator at zero temperature
\cite{Denissenya:2014poa,Denissenya:2014ywa, Denissenya:2015mqa,Denissenya:2015woa}, for a recent theoretical
development see \cite{Lang}. 
The $SU(2)_{CS} \supset U(1)_A$
and $SU(2N_F) \supset SU(N_F)_L \times SU(N_F)_R \times U(1)_A$  symmetries
are symmetries of the chromo-electric interaction in QCD. 
In addition to the chiral transformations the $SU(2)_{CS}$ and $SU(2N_F)$
rotations mix the left- and right-handed components of quark fields.
The chromo-magnetic
interaction as well as the quark kinetic term break these symmetries down
to $SU(N_F)_L \times SU(N_F)_R \times U(1)_A$.

Given the observed new symmetries in high T QCD a question arises whether
the parity doublet model  could still be used as an effective description
for the baryon-like objects at high T. Here we demonstrate that this model
is actually manifestly $SU(2)_{CS}$ and $SU(2N_F)$ invariant.

\section{Chiralspin symmetry \cite{Glozman:2014mka, Glozman:2015qva}}

The  $SU(2)_{CS}$  chiralspin transformations and generators,
defined in the Dirac quark spinor space  are 

\begin{equation}
\label{V-def}
  \psi \rightarrow  \psi^\prime = \exp \left(i  \frac{\varepsilon^n \Sigma^n}{2}\right) \psi  \; ,
\end{equation}
\noindent

\begin{equation}
 \Sigma^n = \{\gamma_k,-i \gamma_5\gamma_k,\gamma_5\},
\label{eq:su2cs_}
\end{equation}
\noindent
$n=1,2,3$. Here $\gamma_k$, $k=1,2,3,4$, is any Hermitian Euclidean gamma-matrix, obeying the following
anticommutation relations

\begin{equation}
\gamma_i\gamma_j + \gamma_j \gamma_i =
2\delta^{ij}; \qquad \gamma_5 = \gamma_1\gamma_2\gamma_3\gamma_4.
\label{eq:diracalgebra}
\end{equation}

\noindent
 Different $k$ define different 
four-dimensional representations that can be reduced into irreducible ones of dim=2.
The $\mathfrak{su}(2)$ algebra
$[\Sigma^a,\Sigma^b]=2i\epsilon^{abc}\Sigma^c$
is satisfied with any Euclidean gamma-matrix.

$U(1)_A$ is a subgroup of $SU(2)_{CS}$.
The  $SU(2)_{CS}$ transformations mix the
left- and right-handed fermions. The free massless quark Lagrangian (1)
does not have this symmetry.

An extension of the direct $SU(2)_{CS} \times SU(N_F)$ product
leads to a $SU(2N_F)$ group. This group contains the chiral
symmetry of QCD $SU(N_F)_L \times SU(N_F)_R \times U(1)_A$ as a subgroup.
Its transformations  are given by
\begin{equation}
\label{W-def}
\psi \rightarrow  \psi^\prime = \exp\left(i \frac{\epsilon^m T^m}{2}\right) \psi\; ,
\end{equation}

\noindent
where $m=1,2,...,(2N_F)^2-1$ and the set of $(2N_F)^2-1$ generators is

\begin{align}
\{
(\tau^a \otimes \mathds{1}_D),
(\mathds{1}_F \otimes \Sigma^n),
(\tau^a \otimes \Sigma^n)
\}
\end{align}
with $\tau$  being the flavour generators with the flavour index $a$ and $n=1,2,3$ is the $SU(2)_{CS}$ index.

The fundamental
vector of $SU(2N_F)$ at $N_F=2$ is

\begin{equation}
\psi =\begin{pmatrix} u_{\textsc{R}} \\ u_{\textsc{L}}  \\ d_{\textsc{R}}  \\ d_{\textsc{L}} \end{pmatrix}. 
\end{equation}
\noindent
The $SU(2N_F)$ transformations mix both flavour and chirality.

While the $SU(2)_{CS}$ and $SU(2N_F)$ symmetries are not symmetries
of the QCD Lagrangian as a whole, they are symmetries of the
fermion charge operator and of the chromo-electric interaction
in QCD. The chromo-magnetic interaction as well as the quark kinetic
term break these symmetries.

\section{Chiralspin symmetry of the parity doublets}

The chiralspin transformation (\ref{V-def}) and generators 
(\ref{eq:su2cs_}) at $k=4$ (the Euclidean $\gamma_4$ matrix coincides with the Minkowskian $\gamma_0$ matrix) can be presented in an  equivalent form, as
follows below.
We will use the chiral representation of the $\gamma$-matrices. Then the
$SU(2)_{CS}$ generators for the representation with $k=4$ are

\begin{equation}
 \Sigma^n = \{ \mathds{1} \otimes \sigma^1,~  \mathds{1} \otimes \sigma^2,
 ~  \mathds{1} \otimes \sigma^3 \}.
\label{eq:su2cseq}
\end{equation}

\noindent
Here $\mathds{1}$ is the unit  $2\times 2$ matrix and the Pauli matrices $\sigma^i$ act in the  space of spinors

\begin{equation}
 \left(\begin{array}{c}
R\\
L
\end{array} \right),
\label{sp}
\end{equation}

\noindent
where R and L  represent the upper and lower components
of the right- and left-handed Dirac bispinors (\ref{RL})

\begin{equation}
\psi_R = \left(\begin{array}{c}
R\\
0
\end{array} \right),
\label{spr}
\end{equation}

\begin{equation}
\psi_L = \left(\begin{array}{c}
0\\
L
\end{array} \right).
\label{spl}
\end{equation}

The $SU(2)_{CS}, ~k=4$ transformation (\ref{V-def}) can then be rewritten as

\begin{equation}
\label{V-defsp}
  \psi \rightarrow  \psi^\prime = \exp \left(i  \frac{\varepsilon^n \Sigma^n}{2}\right) \psi = \exp \left(i  \frac{\varepsilon^n \sigma^n}{2}\right) \left(\begin{array}{c}
R\\
L
\end{array}\right)\; .
\end{equation}
\noindent
The latter equation emphasizes that a reduction of the representations (\ref{V-def}) of $SU(2)_{CS}$ leads to a two-dimensional irreducible
representation.

Now we return to the parity doublet Lagrangian.  Given the Eq. 
(\ref{RLL}) the parity doublet
(\ref{doub}) can be unitary transformed into
a doublet

\begin{equation}
\tilde{\Psi} = \left(\begin{array}{c}
\Psi_R\\
\Psi_L
\end{array} \right),
\label{doubtr}
\end{equation}

\noindent
which is a two-component spinor composed of Dirac bispinors
$\Psi_R$ and $\Psi_L$ (i.e., all together there are  eight components).

The Lagrangian (\ref{lag}) is obviously invariant upon
the $SU(2)_{CS}$ rotation of the doublet (\ref{doubtr}),

\begin{equation}
\left(\begin{array}{c}
\Psi_R\\
\Psi_L
\end{array}\right)\; \rightarrow
\exp \left(i  \frac{\varepsilon^n \sigma^n}{2}\right) \left(\begin{array}{c}
\Psi_R\\
\Psi_L
\end{array}\right)\; .
\label{eq:su2cstra}
\end{equation}

\noindent
It then follows that the parity doublet Lagrangian is not only
chirally invariant under the transformation (\ref{VAD}), but is in addition $SU(2)_{CS}$- and $SU(2N_F)$-invariant with the generators of $SU(2N_F)$
being 

\begin{align}
\{
(\tau^a \otimes \mathds{1}),
(\mathds{1} \otimes \sigma^n),
(\tau^a \otimes \sigma^n)
\}.
\end{align}


\section{Chiralspin symmetric nucleon interpolators}

Here we construct nucleon three-quark interpolators that transform
upon $SU(2)_{CS}$ representations (8-9) with $k=4$. These interpolators should
be used in  lattice studies to establish the chiralspin
symmetry  and consequently to justify  the parity
doublet model as an effective description of the baryon-like objects
at high temperatures.

The quark bispinor, divided into its R- and L-components is

\begin{equation}
q = \left(\begin{matrix}
q^{R}\\
q^{L}
\end{matrix}\right),
\label{eq:quark_field}
\end{equation}

\noindent
where $q = u$ or $d$. It transforms under the representation $\bm{2}$ of $SU(2)_{CS}$. 
Therefore the nucleon three-quark interpolators fall into the representation $\bm{2}\otimes \bm{2}\otimes\bm{2} = \bm{2_1\oplus}\bm{2_2}\oplus\bm{4}$.

Using the standard technique of Young tableaux for representations of the
permutation group  

\begin{equation}
\ytableausetup
  {boxsize=1.20em}
\begin{split}
u_{a_s}^{\alpha}d_{b_j}^{\beta}u_{c_k}^{\gamma} &=
\ytableaushort{\alpha} 
\otimes
\ytableaushort{\beta} 
\otimes
\ytableaushort{\gamma}= \\
&= 
\ytableaushort{\alpha\beta\gamma}\oplus 
\ytableaushort{\alpha\beta,\gamma}\oplus
\ytableaushort{\alpha\gamma,\beta},
\end{split}
\label{eq:young_tab_baryons}
\end{equation}

\noindent
where indices $\alpha,\beta,\gamma$ denote either $L$ or $R$, and e.g. $u_{a_s}^{L}$ is the left-part of the up-quark  with the color index $a$ and the spin index $s$,  we get

\begin{equation}
\ytableausetup
  {boxsize=1.20em}
\begin{array}{llll}
\bm{2_1}: &\ytableaushort{\alpha\gamma,\beta} &\Rightarrow&  \epsilon_{\alpha\beta} 
u_{a_s}^{\alpha}d_{b_j}^{\beta}u_{c_k}^{\gamma}\\
& & &\\
\bm{2_2}: &\ytableaushort{\alpha\beta,\gamma} &\Rightarrow&
u_{a_s}^{\alpha}d_{b_j}^{\beta}u_{c_k}^{\gamma} +
u_{a_s}^{\beta}d_{b_j}^{\alpha}u_{c_k}^{\gamma} +\\
 & &-& u_{a_s}^{\alpha}d_{b_j}^{\gamma}u_{c_k}^{\beta} -
u_{a_s}^{\gamma}d_{b_j}^{\alpha}u_{c_k}^{\beta}\\
\bm{4}: &\ytableaushort{\alpha\beta\gamma} &\Rightarrow& 
u_{a_s}^{\alpha}d_{b_j}^{\beta}u_{c_k}^{\gamma} +
u_{a_s}^{\gamma}d_{b_j}^{\alpha}u_{c_k}^{\beta} + 
u_{a_s}^{\beta}d_{b_j}^{\gamma}u_{c_k}^{\alpha}.\\
\end{array}
\label{eq:young_tab_baryon2}
\end{equation}

Hence the nucleon fields in different irreducible representation of $SU(2)_{CS}$ are given by

\begin{equation}
\bm{2}_1 : \left\lbrace 
\begin{array}{ll}
B_{2_1}(-1/2) = \frac{1}{\sqrt{2}}\left(
u_{a_s}^{L} d_{b_j}^{R} - u_{a_s}^{R} d_{b_j}^{L}
\right)u_{c_k}^{L} \epsilon_{abc}f(s,j,k),\\
\\
B_{2_1}(1/2) = \frac{1}{\sqrt{2}}\left(
u_{a_s}^{L} d_{b_j}^{R} - u_{a_s}^{R} d_{b_j}^{L}
\right)u_{c_k}^{R} \epsilon_{abc}f(s,j,k),
\end{array}\right.
\label{eq:rep_21_b}
\end{equation}

\begin{equation}
\bm{2}_2 : \left\lbrace 
\begin{array}{ll}
B_{2_2}(-1/2) &= \left(
\sqrt{\frac{2}{3}}\:
u_{a_s}^{L} d_{b_j}^{L} u_{c_k}^{R} - \sqrt{\frac{1}{6}}\:
u_{a_s}^{L} d_{b_j}^{R} u_{c_k}^{L}+\right. \\
 & \left. - \sqrt{\frac{1}{6}}\:
u_{a_s}^{R} d_{b_j}^{L} u_{c_k}^{L} \right)
\epsilon_{abc}f(s,j,k),\\
\\
B_{2_2}(1/2) &= \left(
\sqrt{\frac{1}{6}}\:
u_{a_s}^{L} d_{b_j}^{R} u_{c_k}^{R} + \sqrt{\frac{1}{6}}\:
u_{a_s}^{R} d_{b_j}^{L} u_{c_k}^{R}+\right.\\
 & \left. - \sqrt{\frac{2}{3}}\:
u_{a_s}^{R} d_{b_j}^{R} u_{c_k}^{L} \right)
\epsilon_{abc}f(s,j,k),
\end{array}\right.
\label{eq:rep_22_b}
\end{equation}

\begin{equation}
\bm{4} : \left\lbrace 
\begin{array}{ll}
B_{4}(-3/2) &=  u_{a_s}^{L} d_{b_j}^{L} u_{c_k}^{L}\:\epsilon_{abc}f(s,j,k),\\
\\
B_{4}(-1/2) &= \sqrt{\frac{1}{3}}\left(
u_{a_s}^{L} d_{b_j}^{L} u_{c_k}^{R}+
u_{a_s}^{L} d_{b_j}^{R} u_{c_k}^{L}+\right. \\ 
 &+\left. u_{a_s}^{R} d_{b_j}^{L} u_{c_k}^{L}
\right)\:
\epsilon_{abc}f(s,j,k),\\
\\
B_{4}(1/2) &= \sqrt{\frac{1}{3}}\left(
u_{a_s}^{R} d_{b_j}^{R} u_{c_k}^{L} +
u_{a_s}^{R} d_{b_j}^{L} u_{c_k}^{R} +\right.\\
 &+\left. u_{a_s}^{L} d_{b_j}^{R} u_{c_k}^{R}
 \right)\:
\epsilon_{abc}f(s,j,k),\\
\\
B_{4}(3/2) &= u_{a_s}^{R} d_{b_j}^{R} u_{c_k}^{R}\epsilon_{abc}f(s,j,k),
\end{array}\right.
\label{eq:rep_4_b}
\end{equation}

\noindent
where $\epsilon_{abc}$ is the Levi-Civita tensor in the color indices. 
$f(s,j,k)$ is a function of the spin indices; but they are connected with the total angular momentum of the nucleon and at this step we do not perform any summation over $s,j$ and $k$.  
We have denoted with $B_r (\chi_z)$ the baryon field in the $SU(2)_{CS}$
irreducible representation of dim $r = 2\chi+1$   with the chiralspin 
$\chi$
and its projection $\chi_z$.

Now we apply the Eqs. (\ref{eq:rep_21_b}), (\ref{eq:rep_22_b}) and (\ref{eq:rep_4_b}) to nucleon interpolators with isospin $1/2$, total angular momentum  $J = 1/2$ with spin zero diquark.  
The structure of these nucleon interpolators  is \cite{Denissenya:2015woa}:

\begin{equation}
N^{(i)}_{\pm} = \epsilon_{abc} \mathcal{P}_{\pm} \Gamma_1^{(i)} u_a \left(  u_b^T \Gamma^{(i)}_2 d_c  -  d_b^T \Gamma^{(i)}_2 u_c
\right) ,
\label{eq:nucl_1}
\end{equation}

\noindent
where $\mathcal{P}_{\pm} =\frac{1}{2}(\mathds{1} \pm \gamma_4)$.  
The  Dirac structures of $\Gamma_1^{(i)}$ and $\Gamma_2^{(i)}$ are reported in Table \ref{tab:BI}. 

\begin{table}[thb]
\caption{List of Dirac structures for the $N$ baryon fields
with scalar or pseudoscalar diquarks, where $I$ is the isospin, $J^P$ indicates spin and parity.}
\begin{ruledtabular}
\begin{tabular}{c|ccc}
  		$I,J^P$ & $\Gamma_1^{(i)}$ & $\Gamma_2^{(i)}$ & $i$\\
  		\colrule 
  		& $\mathds{1}$ & $C\gamma_5$ & 1 \\
  		$N^{(i)}\left( \frac{1}{2},\frac{1}{2}^{\pm} \right) $ & $\gamma_5$ & $C$ & 2 \\
  		& $i\mathds{1}$ & $C\gamma_5\gamma_4$ & 3\\
 	    & $i\gamma_5$ & $C\gamma_4$ & 4\\
  	\end{tabular}\label{tab:BI}
  	\end{ruledtabular}
\end{table}

Now if we want to write the $N_{\pm}^{(i)}$s as linear combination of the $B_r (\chi_z)$ we need to find the proper expression for $f(s,j,k)$ in the Eqs. (\ref{eq:rep_21_b}), (\ref{eq:rep_22_b}) and (\ref{eq:rep_4_b}). 
At first we note that the diquark  in Eq. (\ref{eq:nucl_1}) has spin zero and it does not carry any spin-index. 
Conversely since $N_{\pm}^{(i)}$ has spin $1/2$, then it carries a spin index given by the quark $u_a$.
This means that in the Eqs. (\ref{eq:rep_21_b}), (\ref{eq:rep_22_b}) and (\ref{eq:rep_4_b}) we need to sum over $j$ and $k$, but not over $s$.
 $C = i\gamma_2 \gamma_4 = \sigma_2 \otimes \sigma_3$, where the Pauli matrix $\sigma_2$ acts in the quark spin space. 
Therefore we can  replace $f(s,j,k)\rightarrow (\sigma_2)_{jk}$ in the Eqs. (\ref{eq:rep_21_b}), (\ref{eq:rep_22_b}) and (\ref{eq:rep_4_b}) and sum over $j$ and $k$ but not over $s$. 
In this case $B_r (\chi_z)$ will represent baryon interpolators with spin $1/2$, in which the diquark  has spin zero. 

Finally we can rewrite the nucleon fields in the Eq. (\ref{eq:nucl_1}) for each $i$ as 

\begin{equation}
\begin{array}{ll}
N_{\pm}^{(1)} &= -\mathcal{P}_{\pm} [
\left(B_4 (-3/2) \pm B_4 (3/2) \right)+\\
&+\sqrt{\frac{1}{6}}\left(B_{2_2}(1/2) \mp  B_{2_2}(-1/2)\right)+ \\
&+\sqrt{\frac{1}{2}}\left(B_{2_1}(1/2) \mp  B_{2_1}(-1/2)\right) +\\
&+\sqrt{\frac{1}{3}}\left( B_{4}(1/2) \pm B_{4}(-1/2)\right)],\\
N_{\pm}^{(2)} &= \mathcal{P}_{\pm} [
-\left(B_4 (-3/2) \pm B_4 (3/2) \right)+\\
&+\sqrt{\frac{1}{6}}\left(B_{2_2}(1/2) \mp  B_{2_2}(-1/2)\right) \\
&+\sqrt{\frac{1}{3}}\left( B_{4} (1/2) \pm B_{4}(-1/2)\right) +\\
&+\sqrt{\frac{1}{2}}\left(  B_{2_1}(1/2) \mp  B_{2_1}(-1/2)\right)],\\
N_{\pm}^{(3)} &= -i\mathcal{P}_{\pm} [
\sqrt{\frac{1}{6}}\left( B_{2_2}(-1/2) \mp  B_{2_2}(1/2)\right) + \\
&+\frac{2}{\sqrt{3}}\left(B_{4}(-1/2) \pm B_{4} (1/2)\right)+\\
&+\sqrt{\frac{1}{2}}\left(  B_{2_1}(-1/2) \mp  B_{2_1}(1/2)\right)],\\
N_{\pm}^{(4)} &= i\mathcal{P}_{\pm} [
-\sqrt{\frac{3}{2}}\left( B_{2_2}(-1/2) \mp  B_{2_2}(1/2)\right)+\\
&+\sqrt{\frac{1}{2}}\left(  B_{2_1}(-1/2) \mp  B_{2_1}(1/2)\right) ],\\
\end{array}
\label{eq:bar_irr1}
\end{equation}

where $B_r (\chi_z)$ are 

\begin{equation}
\begin{array}{ll}
B_{2_1}(-1/2) &= \epsilon_{abc}\frac{1}{\sqrt{2}}\left(\frac{\mathds{1} - \gamma_5}{2}\right)\left[ \gamma_4 u_a \left\lbrace d_b^{T} C \left(\frac{\mathds{1} - \gamma_5}{2}\right)u_c \right\rbrace +\right.\\
&\left. +u_a \left\lbrace d_b^T  C\gamma_4 \left(\frac{\mathds{1} - \gamma_5}{2}\right)u_c \right\rbrace \right],\\
B_{2_1}(1/2) &=\epsilon_{abc}\frac{1}{\sqrt{2}}\left(\frac{\mathds{1} - \gamma_5}{2}\right)\left[ 
u_a \lbrace d_b^T C \left(\frac{\mathds{1} + \gamma_5}{2}\right)u_c \rbrace + \right. \\
&\left.+\gamma_4 u_a \lbrace d_b^T C \gamma_4 \left(\frac{\mathds{1} + \gamma_5}{2}\right)u_c \rbrace \right],\\
B_{2_2}(-1/2) &= \epsilon_{abc}\left(\frac{\mathds{1} - \gamma_5}{2}\right) \left[ -\sqrt{\frac{2}{3}} u_a \lbrace d_b^T  C\gamma_4\left(\frac{\mathds{1} + \gamma_5}{2}\right)u_c\rbrace +\right.\\
&-\sqrt{\frac{1}{6}} u_a \lbrace d_b^T C\gamma_4 \left(\frac{\mathds{1} - \gamma_5}{2}\right) u_c \rbrace +\\
&\left.+\sqrt{\frac{1}{6}}\gamma_4 u_a \lbrace d_b^T C \left(\frac{\mathds{1} - \gamma_5}{2}\right) u_c \rbrace \right],\\
B_{2_2}(1/2) &= \epsilon_{abc}\left(\frac{\mathds{1} - \gamma_5}{2}\right) \left[ 
\sqrt{\frac{1}{6}} u_a \lbrace d_b^T C \left(\frac{\mathds{1} + \gamma_5}{2}\right) u_c \rbrace+\right.\\
&-\sqrt{\frac{1}{6}}\gamma_4 u_a \lbrace d_b^T C \gamma_4 
\left(\frac{\mathds{1} + \gamma_5}{2}\right) u_c \rbrace+ \\
&\left.-\sqrt{\frac{2}{3}}\gamma_4 u_a \lbrace d_b^T C \gamma_4\left(\frac{\mathds{1} - \gamma_5}{2}\right) u_c\rbrace \right],\\
B_{4}(-3/2) &= -\epsilon_{abc}\left(\frac{\mathds{1} - \gamma_5}{2}\right)  u_a \lbrace d_b^{T} C \left(\frac{\mathds{1} - \gamma_5}{2}\right) u_c\rbrace, \\
B_{4} (-1/2) &= \sqrt{\frac{1}{3}}\epsilon_{abc}\left(\frac{\mathds{1} - \gamma_5}{2}\right) \left[ -u_a \lbrace d_b^T C\gamma_4 \left(\frac{\mathds{1} + \gamma_5}{2}\right)u_c \rbrace + \right.\\
&+ u_a \lbrace d_b^T C\gamma_4 \left(\frac{\mathds{1} - \gamma_5}{2}\right)u_c \rbrace+\\
&\left.-\gamma_4 u_a \lbrace d_b^T C \left(\frac{\mathds{1} - \gamma_5}{2}\right)u_c \rbrace\right],\\
B_{4} (1/2) &= \sqrt{\frac{1}{3}}\epsilon_{abc}\left(\frac{\mathds{1} - \gamma_5}{2}\right) \left[ \gamma_4 u_a \lbrace d_b^T  C\gamma_4 \left(\frac{\mathds{1} - \gamma_5}{2}\right)u_c \rbrace + \right.\\
&- \gamma_4 u_a \lbrace d_b^T C\gamma_4 \left(\frac{\mathds{1} + \gamma_5}{2}\right)u_c \rbrace+\\
&\left.+ u_a \lbrace d_b^T C \left(\frac{\mathds{1} + \gamma_5}{2}\right)u_c \rbrace\right],\\
B_{4}(3/2) &= \epsilon_{abc}\left(\frac{\mathds{1} - \gamma_5}{2}\right) \gamma_4 u_a \lbrace d_b^T C \left(\frac{\mathds{1} + \gamma_5}{2}\right) u_c \rbrace, \\
\end{array}
\label{eq:rep_nucl}
\end{equation}

\noindent
and the brace brackets mean antisymmetrization in the flavour space,
 e.g. $\lbrace d_b^{T} C u_c \rbrace = d_b^{T} C u_c - u_b^{T} C d_c$.
The parity projector $\mathcal{P}_{\pm}$ in Eq. (\ref{eq:bar_irr1}) can be replaced with $\mathcal{P}_{\pm} \rightarrow \mathds{1}/\sqrt{2}$, since in the computation of the correlators the presence of $\mathcal{P}_{\pm}$ is irrelevant.

Hence we have rewritten  the nucleon interpolators   $N_{\pm}^{(i)}$  in terms of interpolators that transform according to the irreducible representations of $SU(2)_{CS}$.

The cross-correlation matrix for interpolators  $B_{r}(\chi_z)$ 
and $B_{r\prime}({\chi\prime}_z)$ is

\begin{equation}
C (t)_{\chi,\chi\prime,\chi_z, {\chi\prime}_z} = \langle 0 \vert B_r(\chi_z;t)B_{r\prime}( {\chi\prime}_z;0)^{\dagger}\vert 0 \rangle.
\end{equation}

\noindent
The $SU(2)_{CS}$ restoration requires that the cross-correlators of operators
from different
representations of $SU(2)_{CS}$ vanish and the cross-correlators of operators  within 
a given representation of $SU(2)_{CS}$ that are diagonal in indices
 $\chi_z$ and  ${\chi\prime}_z$ coincide while the off-diagonal vanish.
In this case the diagonal correlators $C(t)_{N_{\pm}^{(i)}}$ of nucleon interpolators $N_{\pm}^{(i)}$
from the Table 1 are given as

\begin{equation}
\begin{array}{lll}
C(t)_{N_{\pm}^{(i)}} &= \frac{2}{3}C(t)_{3/2} + \frac{1}{12}C(t)_{1/2_2} +& \frac{1}{4}C(t)_{1/2_1}\\
 & & \mbox{for}\;i=1,2,3\\
\\
C(t)_{N_{\pm}^{(4)}} &= \frac{1}{4}C(t)_{1/2_2} + \frac{3}{4}C(t)_{1/2_1}.
\end{array}
\label{eq:corr_nucl}
\end{equation}

\noindent
Here $C(t)_{3/2}$ is a correlator $C (t)_{\chi,\chi\prime,\chi_z, {\chi\prime}_z}$ with $\chi=\chi\prime = 3/2$ and with any $\chi_z = {\chi\prime}_z$, and similar for $C(t)_{1/2_1}$ and $C(t)_{1/2_2}$.
Hence in the $SU(2)_{CS}$-symmetric regime
 all correlators $C(t)_{N_{\pm}^{(i)}}$ with $i=1,2,3$ should be degenerate. Such a degeneracy
was indeed observed at zero temperature upon truncation of the near-zero modes of the Dirac
operator in ref. \cite{Denissenya:2015woa}.

\section{Conclusions}

In this paper we have demonstrated that a chirally symmetric
parity doublet Lagrangian is not only chirally symmetric, but in addition
is $SU(2)_{CS}$ and $SU(2N_F)$ invariant. Then we have constructed
nucleon three-quark interpolators that transform under irreducible
representations of $SU(2)_{CS}$. Such interpolators are required for
lattice studies to establish the chiralspin symmetry.

\begin{acknowledgments} 
This work is supported by the Austrian Science Fund FWF through grants  
DK W1203-N16 and P26627-N27. 
The authors thank Christian Rohrhofer for
cross-checking of all equations of section 4.
\end{acknowledgments}


\appendix

\section{}\label{app:A}

We begin with the Lagrangian (2.36) of \cite{TIT} with two Dirac fermions

\begin{equation}
\mathcal{L} = i\bar{\psi}_1 \gamma_{\mu}\partial^{\mu}\psi_1 + 
i\bar{\psi}_2 \gamma_{\mu}\partial^{\mu}\psi_2 
-m(\bar{\psi}_1\psi_2 + \bar{\psi}_2 \psi_1),
\label{eq:mirror_lagr}
\end{equation}

\noindent
which is invariant under chiral transformation with the ``mirror assignment":

\begin{equation}
\psi_1 \rightarrow \exp\left(i\alpha\gamma_5\right)\psi_1 ,\qquad \psi_2\rightarrow \exp\left(-i\alpha\gamma_5\right) \psi_2.
\label{eq:mirror_trans}
\end{equation}

\noindent
From this we can obtain the chiral transformation of the doublet (\ref{doub})

\begin{equation}
\Psi = \left(\begin{matrix}
\Psi_+\\
\Psi_-
\end{matrix}\right),
\label{eq:doub2}
\end{equation}

\noindent
where 

\begin{equation}
\begin{split}
&\Psi_+ = \frac{1}{\sqrt{2}}(\psi_1 + \psi_2),\\
&\Psi_- = \frac{1}{\sqrt{2}}(\gamma_5\psi_1 -\gamma_5 \psi_2).
\end{split}
\end{equation}

\noindent
Then upon the ``mirror" transformation (\ref{eq:mirror_trans}) the parity doublet (\ref{eq:doub2}) transforms as:

\begin{equation}
\left(\begin{matrix}
\Psi_+\\
\Psi_-
\end{matrix}\right)
\rightarrow \exp\left( i\alpha\sigma^1\right) 
\left(\begin{matrix}
\Psi_+\\
\Psi_-
\end{matrix}\right).
\end{equation}

\noindent
Extending the $U(1)_A$ mirror transformation (\ref{eq:mirror_trans}) 
to the axial part of the mirror $SU(2)_L \times SU(2)_R$ transformation

\begin{equation}
\begin{split}
&\psi_1 \rightarrow \exp\left(i \frac{\alpha^a \tau^a}{2}\gamma^5\right) \psi_1,\\
&\psi_2 \rightarrow \exp\left(-i\frac{\alpha^a \tau^a}{2}\gamma^5 \right) \psi_2,
\end{split}
\end{equation}

\noindent
we obtain the axial part of the $SU(2)_L \times SU(2)_R$ transformation of the parity doublet (\ref{VAD}):

\begin{equation}
\left(\begin{matrix}
\Psi_+\\
\Psi_-
\end{matrix}\right)
\rightarrow \exp\left(i\frac{\alpha^a \tau^a}{2}\otimes\sigma^1\right) 
\left(\begin{matrix}
\Psi_+\\
\Psi_-
\end{matrix}\right).
\end{equation}

\end{document}